\documentclass{appolb}
\usepackage{epsfig}
\usepackage{graphics, color, cite, curves, amssymb, amsmath, rotating, epic}
\usepackage{eepic}

\newcommand{\gaz}{g_A^{\mbox{$\scriptscriptstyle (Z)$}}}
\newcommand{\run}[1]{\widetilde{\alpha}_{#1}}
\newcommand{\hsp}[1]{\hspace*{#1 mm}}
\newcommand{\smallfrac}[2]{\mbox{\small ${\displaystyle \frac{#1}{#2}}$}}
\newcommand{\footfrac}[2]%

\newcommand{\cxx}{1 - {{4x^{2} P^{2}}\over{Q^{2}}} }
\def\cx{1 - {{2x    P^{2}}\over{Q^{2}}} }

\def\cut{\sqrt{1 - {{4(m^{2}+\lambda^2)}\over {s}} } }

\begin{document}

\title{
FIXED POLES, POLARIZED GLUE AND NUCLEON SPIN STRUCTURE
\thanks{Presented at the 43rd Cracow School of Theoretical Physics:
Fundamental Interactions, Zakopane, Poland May 30 - June 8, 2003.}
}
\author{Steven D. Bass
\address{
High Energy Physics Group, \\
Institute for Experimental Physics
and Institute for Theoretical Physics, \\
Universit\"at Innsbruck, \\
Technikerstrasse 25, A 6020 Innsbruck, Austria}
}
\maketitle
\begin{abstract}
We review the theory and present status of the proton spin problem with
emphasis on possible gluonic and sea contributions.
We discuss the possibility of a $J=1$ fixed pole correction
to the
Ellis-Jaffe sum rule for polarized deep inelastic scattering.
Fixed poles in the real part of the forward Compton scattering
amplitude
have the potential to induce subtraction constant corrections
to sum rules for photon nucleon scattering.
\end{abstract}

\section{Introduction}

Understanding the spin structure of the proton is one of the most 
challenging problems facing subatomic physics:
How is the spin of the proton built up out from the intrinsic spin
and orbital angular momentum of its quark and gluonic constituents ?
What happens to spin in the transition between current and constituent
quarks in low-energy QCD.
Key issues include the role of polarized glue and gluon topology in 
building up the spin of the proton.

Our present knowledge about the spin structure of the nucleon comes from
polarized deep inelastic scattering.
Following pioneering experiments at SLAC \cite{slac}, recent experiments
in fully inclusive polarized deep inelastic scattering have extended
measurements of the nucleon's $g_1$ spin dependent structure function
to lower values of Bjorken $x$ where the nucleon's sea becomes important
\cite{windm}.
From the first moment of $g_1$, 
these experiments have 
been interpreted to imply a small value for the flavour-singlet axial-charge:
\begin{equation}
g_A^{(0)}\bigr|_{\rm pDIS} = 0.2 - 0.35 .
\label{eqa1}
\end{equation}
This result is particularly interesting \cite{spinrev,bass99} 
because $g_A^{(0)}$ is
interpreted in the parton model as the fraction of the proton's spin which is
carried
by the intrinsic spin of its quark and antiquark constituents.
The value (\ref{eqa1}) is about half the prediction of
relativistic constituent quark models ($\sim 60\%$).
It corresponds to a negative strange-quark polarization
\begin{equation}
\Delta s = -0.10 \pm 0.04 
\label{eqa2}
\end{equation}
(polarized in the opposite direction to the spin of the proton).

The small value of $g_A^{(0)}|_{\rm pDIS}$ 
extracted from 
polarized deep inelastic scattering has inspired vast 
experimental and theoretical activity to understand the spin
structure of the proton. New experiments are underway or being planned
to map out the proton's spin-flavour structure and to measure the amount of
spin carried by polarized gluons in the polarized proton.
These include semi-inclusive polarized deep inelastic scattering, 
polarized proton-proton collisions at RHIC \cite{rhic}, 
and polarized $ep$ collider studies \cite{bassdr}.
Experiments at JLab will map out the valence region 
at large Bjorken $x$ (close to one) \cite{meziani}.
An independent, weak interaction, measurement of 
$g_A^{(0)}$ could be performed using elastic neutrino proton scattering
\cite{tayloe}.
Experiments with transversely polarized targets are just beginning and 
promise to reveal new information about the spin structure of the proton
including tests of the Burkhardt-Cottingham sum rule for the nucleon's 
$g_2$ spin structure function and measurements of a whole new family of 
``transversity observables''.

The plan of these lectures is as follows.
We first summarise the phenomenology of the proton spin problem, including 
possible gluonic contributions.
Next, in Sections 2 and 3, we give an overview of the derivation of 
the spin sum rules for polarized photon nucleon scattering, detailing 
the assumptions that are made at each step.
Here we explain how these sum rules could be affected by potential 
subtraction constants (subtractions at infinity) in the dispersion 
relations for the spin dependent part of the forward Compton amplitude.
We next give a brief review of fixed pole contributions to deep inelastic 
scattering in Section 4.
Fixed poles are well known to play a vital role in the Adler sum rule 
for W-boson nucleon scattering \cite{adler} and the Schwinger term sum rule 
for the longitudinal structure function measured in unpolarized deep 
inelastic $ep$ scattering \cite{bgj}.
We explain how fixed poles could, in principle, affect the sum rules 
for the first moments of the $g_1$ and $g_2$ spin structure functions.
For example, a subtraction constant correction to the Ellis-Jaffe 
sum rule for the first moment of the nucleon's $g_1$ spin dependent 
structure function would follow if there is a real constant term in 
the spin dependent part of the forward deeply virtual Compton scattering 
amplitude.
Section 5 discusses the QCD axial anomaly and its manifestation in 
$g_A^{(0)}$ and the spin structure of the proton. 
We conjecture that gluon topology may induce a $J=1$ 
fixed pole correction to the Ellis-Jaffe sum rule.
Photon-gluon fusion and its importance to semi-inclusive measurements of 
sea polarization in polarized deep inelastic scattering are disussed in 
Section 6.
A summary of key issues is given in Section 7.

\subsection{The proton spin problem}

First consider the flavour-singlet channel.

In QCD the axial anomaly \cite{anomaly} induces gluonic contributions 
to the flavour-singlet axial charge associated with the polarized glue 
in the nucleon and with gluon topology.
\begin{enumerate}
\item
The first moment of the $g_1$ spin structure function for polarized
photon-gluon fusion $(\gamma^* g \rightarrow q {\bar q})$
receives a negative contribution $-{\alpha_s \over 2 \pi}$
from $k_t^2 \sim Q^2$, 
where
$k_t$ is the quark transverse momentum 
relative to the photon gluon direction and
$Q^2$ is the virtuality of the hard photon \cite{ccm,bint}.
It also receives a
positive contribution 
(proportional to the mass squared of the struck quark or antiquark) 
from low values of
$k_t$, 
$k_t^2 \sim P^2, m^2$ where $P^2$ is the virtuality of the parent
gluon and
$m$ is the mass of the struck quark.
The contact interaction ($k_t \sim Q$) between the polarized photon and 
gluon is flavour-independent, associated with the QCD axial anomaly and 
measures the spin of the target gluon.
The mass dependent contribution is absorbed into the quark wavefunction 
of the nucleon. 
\item
Gluon topology is associated with gluonic boundary conditions and 
has the potential to induce a topological contribution 
to $g_A^{(0)}$ associated with Bjorken $x$ equal to zero:
topological $x=0$ polarization or, 
essentially, a spin polarized condensate inside a nucleon \cite{topology}.
\end{enumerate}
Putting this physics together leads to the formula 
\cite{ccm,bint,topology,etar}:
\begin{equation}
g_A^{(0)}
=
\biggl(
\sum_q \Delta q - 3 {\alpha_s \over 2 \pi} \Delta g \biggr)_{\rm partons}
+ {\cal C}
\label{eqa3}
\end{equation}
Here $\Delta g_{\rm partons}$ is the amount of spin carried by polarized 
gluon partons in the polarized proton and 
$\Delta q_{\rm partons}$ measures the spin carried by quarks and 
antiquarks 
carrying ``soft'' transverse momentum $k_t^2 \sim m^2, P^2$;
${\cal C}$ denotes the topological contribution.
Since $\Delta g \sim 1/\alpha_s$ under QCD evolution, the
polarized gluon term $[-{\alpha_s \over 2 \pi} \Delta g]$
in Eq.(\ref{eqa3})
scales as $Q^2 \rightarrow \infty$ \cite{etar}.

Understanding the transverse momentum dependence of the quark and gluon 
contributions in Eq.(\ref{eqa3}) is essential to ensure that theory and 
experimental acceptance are correctly matched when extracting information 
from semi-inclusive measurements aimed at disentangling the individual 
valence, sea and gluonic contributions \cite{bass03}.

Since $x=0$ is inaccessible to deep inelastic scattering, 
the deep inelastic measurement of $g_A^{(0)}$, Eq.(\ref{eqa1}),
is not necessarily inconsistent with the constituent quark model 
prediction 0.6
{\it if} a substantial fraction of the spin of the constituent quark 
is associated with gluon topology in the transition from constituent
to current quarks  (measured in polarized deep inelastic scattering)
through dynamical axial U(1) symmetry breaking \cite{bass99}.

An ``$x=0$'' correction to deep inelastic measurements of $g_A^{(0)}$
would also follow if there is a leading twist 
``subtraction at infinity'' in the dispersion relation for the spin 
dependent part of the forward Compton scattering amplitude  
(from a $J=1$ Regge fixed pole).
An independent measurement of the flavour-singlet axial-charge
through elastic neutrino proton scattering would be extremely valuable.

\subsection{The isovector part of $g_1$}

Quark model predictions for $g_1$ work much better in the isovector channel. 
The Bjorken sum rule which relates the first moment of $(g_1^p - g_1^n)$ 
to the isovector axial charge $g_A^{(3)}$ measured in neutron beta decays
has been confirmed at the level of 10\% \cite{windm}.

Looking beyond the first moment, the shape of $(g_1^p - g_1^n)$ 
is very interesting.
Figure 1 from Ref.\cite{bass99}
shows $2 x (g_1^p - g_1^n)$ (SLAC data) 
together with the isovector structure function $(F_2^p - F_2^n)$ (NMC data).
The ratio
$R_{(3)} =
 2 x (g_1^p - g_1^n) / (F_2^p - F_2^n)$
is plotted in Fig.2.
It measures the ratio of polarized to unpolarized isovector quark 
distributions.
The ratio $R_{(3)}$ is observed to be approximately constant 
(at the value $\sim 5/3$ 
predicted by SU(6) constituent quark models) 
for $x$ between 0.03 and 0.2,
and goes towards one when $x \rightarrow 1$
(consistent with the predictions of QCD counting rules \cite{brodbs}).
The area under $(F_2^p-F_2^n)/2x$ is fixed by the Gottfried integral
\cite{gottfried}.
The observed shape of
$g_1^p - g_1^ n$
is almost {\it required} \cite{bass99}
in order to reproduce the area under the Bjorken sum rule, 
which is fixed by the value of $g_A^{(3)}$.
The constant ratio in the low to medium $x$ range contrasts 
with the naive Regge prediction (strictly for $Q^2=0$)
that the ratio $R_{(3)}$ should be roughly proportional to $x$ 
as
$x \rightarrow 0$.
\begin{figure}[h] 
\includegraphics{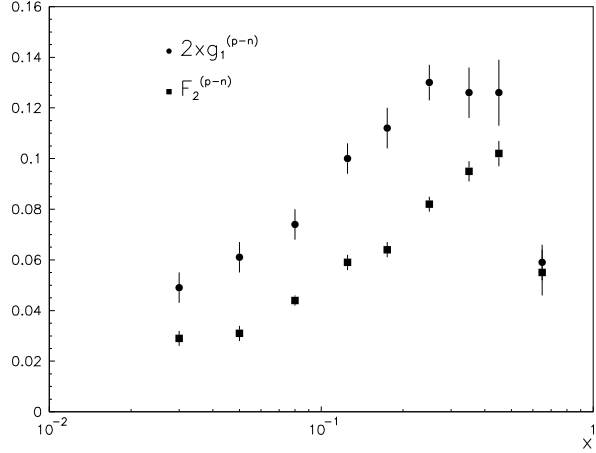} 
\begin{center} 
\vspace{6.5cm} 
\parbox{12.0cm} 
{\caption[Delta]
{
The isovector structure functions $2x g_1^{(p-n)}$ (SLAC data) and 
$F_2^{(p-n)}$ (NMC).
}
\label{fig1}} 
\end{center} 
\end{figure}
\begin{figure}[ht] 
\includegraphics{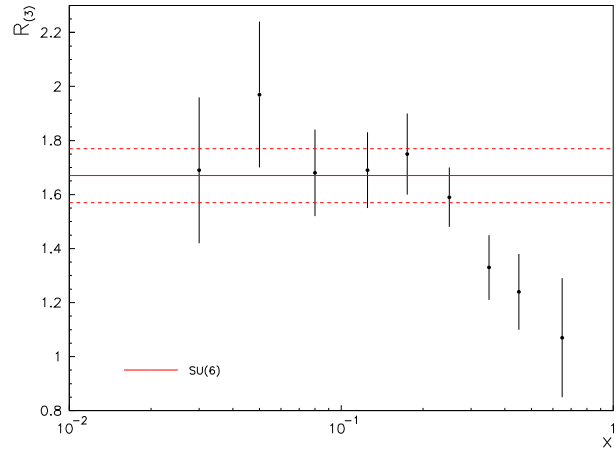}
\begin{center} 
\vspace{6.5cm} 
\parbox{12cm} 
{\caption[Delta]
{ 
The ratio $R_{(3)} = 2x g_1^{(p-n)}/F_2^{(p-n)}$.
}
\label{fig2}} 
\end{center} 
\end{figure}

\section{Scattering amplitudes and cross-sections}

The spin dependent structure functions $g_1$ and $g_2$ 
are defined through 
the imaginary part of the forward Compton scattering amplitude.
Consider the amplitude for forward scattering of a photon 
carrying momentum $q_{\mu}$ ($q^2 = -Q^2 \leq 0$) 
from a polarized nucleon 
with momentum $p_{\mu}$, mass $M$ and spin $s_{\mu}$.
Let $J_{\mu}(z)$ denote the electromagnetic current in QCD.
The forward Compton amplitude
\begin{equation}
T_{\mu \nu}(q,p) =
i \int d^{4}z \ e^{iq.z}
               \langle p, s | \ T(J_{\mu}(z) J_{\nu}(0) ) \ | p, s \rangle
\label{eqba}
\end{equation}
is given by the sum of spin independent 
(symmetric in $\mu$ and $\nu$) 
and
spin dependent (antisymmetric in $\mu$ and $\nu$)
contributions:
\begin{eqnarray}
T_{\mu \nu}^{S} 
&=&
{1 \over 2} (T_{\mu \nu} + T_{\nu \mu})
\nonumber \\
&=&
- T_1 (g_{\mu \nu} + {q_{\mu} q_{\nu} \over Q^{2} })
             + {1 \over M^2} T_2 
             (p_{\mu} + {p.q \over Q^{2}} q_{\mu})
             (p_{\nu} + {p.q \over Q^{2}} q_{\nu})
\label{eqbb}
\end{eqnarray}
and
\begin{eqnarray}
T_{\mu \nu}^{A} 
&=&
{1 \over 2} (T_{\mu \nu} - T_{\nu \mu})
\nonumber \\
&=&
{i \over M^2} 
\epsilon_{\mu \nu \lambda \sigma} 
q^{\lambda} 
\biggl[ 
s^{\sigma}
( A_1 + {\nu \over M} A_2 ) - {1 \over M^2} s.q p^{\sigma} A_2 \biggr]
\label{eqbc}
\end{eqnarray}
Here $\nu = p.q / M$,
$\epsilon_{0123} =+1$,
and
the proton spin vector is normalized to $s^2 = -1$.
The form-factors $T_1$, $T_2$, $A_1$ and $A_2$ are functions of $\nu$ 
and $Q^2$.

The hadron tensor for inclusive photon nucleon scattering, which contains 
the spin dependent structure functions, 
is obtained from the imaginary part of $T_{\mu \nu}$:
\begin{equation}
W_{\mu \nu} = {1 \over \pi} {\rm Im} T_{\mu \nu}
= {1 \over 2 \pi} 
     \int d^{4}z \ e^{iq.z} 
\langle p, s | \ [J_{\mu}(z), J_{\nu}(0)] \ |p, s \rangle 
\label{eqbd}
\end{equation}
Here the connected matrix element is understood
(indicating that the photon interacts with the target and not the vaccum).
The spin independent and spin dependent components of $W_{\mu \nu}$ 
are
\begin{equation}
W_{\mu \nu}^{S} = - W_1 (g_{\mu \nu} + {q_{\mu} q_{\nu} \over Q^{2} })
             + {1 \over M^2} W_2 
             (p_{\mu} + {p.q \over Q^{2}} q_{\mu})
             (p_{\nu} + {p.q \over Q^{2}} q_{\nu})
\label{eqbe}
\end{equation}
and
\begin{equation}
W_{\mu \nu}^{A} = 
{i \over M^2} 
\epsilon_{\mu \nu \lambda \sigma} 
q^{\lambda} 
\biggl[ 
s^{\sigma}
( G_1 + {\nu \over M} G_2 ) - {1 \over M^2} s.q p^{\sigma} G_2 \biggr]
\label{eqbf}
\end{equation}
respectively.
The cross sections for the absorption of a transversely polarized 
photon with spin polarized parallel 
$\sigma_{3 \over 2}$ and anti-parallel 
$\sigma_{1 \over 2}$ to the spin of the target nucleon are:
\begin{eqnarray}
\sigma_{3 \over 2} &=& {4 \pi^2 \alpha \over \sqrt{\nu^2 + Q^2}}
\biggl[ W_1 - {\nu \over M^2} G_1 + {Q^2 \over M^3} G_2 \biggr]
\nonumber \\
\sigma_{1 \over 2} &=& {4 \pi^2 \alpha \over \sqrt{\nu^2 + Q^2}}
\biggl[ W_1 + {\nu \over M^2} G_1 - {Q^2 \over M^3} G_2 \biggr]
\label{eqbg}
\end{eqnarray}
where we use usual conventions for the virtual photon flux factor 
\cite{roberts}.
The spin dependent part of the inclusive photon nucleon cross section 
is:
\begin{equation}
\sigma_{1 \over 2} - \sigma_{3 \over 2}
= {8 \pi^2 \alpha \over \sqrt{\nu^2 + Q^2}}
\biggl[ {\nu \over M^2} G_1 - {Q^2 \over M^3} G_2 \biggr]
\label{eqbh}
\end{equation}
For real photons ($Q^2=0$) this equation becomes
$
\{
\sigma_{1 \over 2} - \sigma_{3 \over 2}
= {8 \pi^2 \alpha \over M^2} G_1
\}
$
---
that is, $G_2$ decouples from polarized photoproduction.
The $W_2$ structure function is measured in unpolarized lepton nucleon
scattering through the absorption of longitudinally and transversely
polarized photons.
In high $Q^2$ deep inelastic scattering the structure functions exhibit
approximate  scaling:
\begin{eqnarray}
  M \ W_1 (\nu, Q^2) &\rightarrow& F_1 (x, Q^2)  \nonumber \\
\nu \ W_2 (\nu, Q^2) &\rightarrow& F_2 (x, Q^2) \nonumber \\
{\nu \over M} \ G_1 (\nu, Q^2) &\rightarrow& g_1 (x, Q^2) \nonumber \\
{\nu^2 \over M^2} \ G_2(\nu, Q^2) &\rightarrow& g_2 (x, Q^2)
\label{eqbi}
\end{eqnarray}
Here $x= {Q^2 \over 2m \nu}$ is the Bjorken variable.
The structure functions 
$F_1$, $F_2$, $g_1$ and $g_2$ 
scale 
modulo perturbative QCD logarithmic evolution in $Q^2$.

Regge theory makes predictions for the high-energy asymptotic behaviour of 
the structure functions:
\begin{eqnarray}
&W_1& \sim \nu^{\alpha}   \nonumber \\
&W_2& \sim \nu^{\alpha -2} \nonumber \\
&G_1& \sim \nu^{\alpha -1}  \nonumber \\
&G_2& \sim \nu^{\alpha -1}
\label{eqbl}
\end{eqnarray}
Here $\alpha$ denotes the (effective) intercept for the leading 
Regge exchange contributions.
The Regge predictions for the leading exchanges include 
$\alpha = 1.08$ 
for the pomeron contributions to $W_1$ and $W_2$, 
and $\alpha \simeq 0.5$ for the $\rho$ and $\omega$ 
exchange contributions to the spin independent structure functions.
For $G_1$ the leading gluonic exchange behaves as $\{ \ln \nu \} / \nu$ 
\cite{clos,sbpl}; 
there are also isovector $a_1$ and isoscalar $f_1$ Regge exchanges
\cite{heim}.
If one makes the usual assumption that the $a_1$ and $f_1$ Regge
trajectories are straight lines parallel to the $(\rho, \omega)$ 
trajectories then one finds 
$\alpha_{a_1} \simeq \alpha_{f_1} \simeq -0.4$,
within the phenomenological range $-0.5 \leq \alpha_{a_1} \leq 0$
\cite{ek}.
For $G_2$ one expects contributions from possible multi-pomeron 
(three or more) cuts ($\sim (\ln \nu)^{-5}$) 
and Regge-pomeron cuts ($\sim \nu^{\alpha_i(0)-1} / \ln \nu $) 
with
$\alpha_i(0) <1$ 
(since the pomeron does not couple to $A_1$ or $A_2$) \cite{ioffebook}.
The effective intercepts for small $x$, or high $\nu$, 
physics increase 
with increasing $Q^2$ through perturbative QCD evolution.

\section{Dispersion Relations and Spin Sum Rules}

Sum rules for the (spin) structure functions are derived using 
dispersion relations and, 
for deep inelastic scattering, the operator product expansion.
For fixed $Q^2$ the forward Compton scattering amplitude 
$T_{\mu \nu}(\nu,Q^2)$ 
is analytic in the photon energy $\nu$ except for branch cuts 
along the positive real axis for $|\nu| \geq Q^2/2M$.
Crossing symmetry implies that
\begin{eqnarray}
A_1^* (Q^2, -\nu) &=&   A_1 (Q^2, \nu) \nonumber \\
A_2^* (Q^2, -\nu) &=& - A_2 (Q^2, \nu) 
\label{eqcd}
\end{eqnarray}
The spin structure functions in the imaginary parts of $A_1$ and $A_2$
satisfy the crossing relations
\begin{eqnarray}
G_1 (Q^2, -\nu) &=&  - G_1 (Q^2, \nu) \nonumber \\
G_2 (Q^2, -\nu) &=&  + G_2 (Q^2, \nu) 
\label{eqce}
\end{eqnarray}
For $g_1$ and $g_2$ these relations become
\begin{eqnarray}
g_1 (x, Q^2) &=&  + g_1 (-x, Q^2) \nonumber \\
g_2 (x, Q^2) &=&  + g_2 (-x, Q^2)
\label{eqcf}
\end{eqnarray}
We use Cauchy's integral theorem and the crossing relations to derive
dispersion relations for $A_1$ and $A_2$.
Assuming that the asymptotic behaviour of the spin structure functions
$G_1$ and $G_2$ yield convergent integrals in an unsubtracted dispersion 
relation
we are tempted
to write unsubtracted dispersion relations:
\begin{eqnarray}
A_1 (Q^2, \nu) 
&=& 
{2 \over \pi} \int_{Q^2/2M}^{\infty} \ {\nu' d \nu' \over \nu'^2 - \nu^2}
\ {\rm Im} A_1 (Q^2, \nu')
\nonumber \\
A_2 (Q^2, \nu) 
&=& 
{2 \over \pi} \nu \int_{Q^2/2M}^{\infty} \ {d \nu' \over \nu'^2 - \nu^2}
\ {\rm Im} A_2 (Q^2, \nu')
\label{eqcg}
\end{eqnarray}
These expressions can be rewritten as dispersion relations involving 
$g_1$ and $g_2$.
We define:
\begin{eqnarray}
\alpha_1 (\omega, Q^2) &=& {\nu \over M} \ A_1 \nonumber \\
\alpha_2 (\omega, Q^2) &=& {\nu^2 \over M^2} \ A_2
\label{eqch}
\end{eqnarray}
Then, the formulae in (\ref{eqcg}) become
\begin{eqnarray}
\alpha_1 (\omega, Q^2) 
&=& 
2 \omega 
\int_1^{\infty} \ {d \omega' \over \omega'^2 - \omega^2}
\ g_1 (\omega', Q^2)
\nonumber \\
\alpha_2 (\omega, Q^2) 
&=& 
2 \omega^3 
\int_1^{\infty} \ {d \omega' \over \omega'^2 (\omega'^2 - \omega^2)}
\ g_2 (\omega', Q^2)
\label{eqci}
\end{eqnarray}
where $\omega = {1 \over x} = {2 M \nu \over Q^2}$.

In general there are two alternatives to an unsubtracted dispersion 
relation. 
\begin{enumerate}
\item
First, if the high energy behaviour of $G_1$ and/or $G_2$ (at some fixed
$Q^2$)
produced a divergent integral, then the dispersion relation would require
a subtraction.
Regge predictions for the high energy behaviour of $G_1$ and $G_2$
-- see below Eq.(\ref{eqbl}) -- 
each lead to convergent integrals so this scenario is not expected to occur.
\item
Second, 
even if the integral in the unsubtracted relation converges, there is 
still
the potential for a ``subtraction at infinity''.
This scenario would occur if the real part of $A_1$ and/or $A_2$ 
does not vanish sufficiently fast enough when 
$\nu \rightarrow \infty$ so that we pick up a finite contribution 
from the contour 
(or 
``circle at infinity'').
In the context of Regge theory such subtractions can arise from fixed
poles 
(with 
$J = \alpha(t) = 0$ in $A_2$ 
 or 
$J = \alpha(t) = 1$ in $A_1$
for all $t$) 
in the real part of
the
forward Compton amplitude.
We shall discuss these fixed poles and potential subtractions in Section 4.
\end{enumerate}
In the presence of a potential 
``subtraction at infinity'' 
the dispersion relations (\ref{eqcg}) are modified to:
\begin{eqnarray}
A_1 (Q^2, \nu) 
&=& 
{\cal P}_1 (\nu, Q^2) +
{2 \over \pi} \int_{Q^2/2M}^{\infty} \ {\nu' d \nu' \over \nu'^2 - \nu^2}
\ {\rm Im} A_1 (q^2, \nu')
\nonumber \\
A_2 (Q^2, \nu) 
&=& 
{\cal P}_2 (\nu, Q^2) +
{2 \over \pi} \nu \int_{Q^2/2M}^{\infty} \ {d \nu' \over \nu'^2 - \nu^2}
\ {\rm Im} A_2 (q^2, \nu')
\label{eqcj}
\end{eqnarray}
Here 
\begin{eqnarray}
{\cal P}_1 (\nu, Q^2) &=& \beta_1 (Q^2) 
\nonumber \\
{\cal P}_2 (\nu, Q^2) &=& \beta_2 (Q^2) {M \over \nu}
\label{eqck}
\end{eqnarray}
denote the subtraction constants. 
The crossing relations (\ref{eqcd}) 
for $A_1$ and $A_2$ 
are observed by the functions ${\cal P}_i$.
Scaling requires that
$\beta_1 (Q^2)$ and $\beta_2 (Q^2)$ 
(if finite) must be nonpolynomial in $Q^2$ -- see Section 4.
The equations (\ref{eqcj}) can be rewritten:
\begin{eqnarray}
\alpha_1 (\omega, Q^2) 
&=& 
{Q^2 \over 2 M^2 } \ \beta_1(Q^2) \ \omega  +
2 \omega 
\int_1^{\infty} \ {d \omega' \over \omega'^2 - \omega^2}
\ g_1 (\omega', Q^2)
\nonumber \\
\alpha_2 (\omega, Q^2) 
&=& 
{Q^2 \over 2 M^2 } \ \beta_2 (Q^2) \ \omega +
2 \omega^3 
\int_1^{\infty} \ {d \omega' \over \omega'^2 (\omega'^2 - \omega^2)}
\ g_2 (\omega', Q^2)
\label{eqcl}
\end{eqnarray}
Next, the fact that both $\alpha_1$ and $\alpha_2$ are analytic 
for $| \omega | \leq 1$ allows us to make the Taylor series expansions
(about $\omega = 0$):
\begin{eqnarray}
\alpha_1 (x, Q^2) 
&=& 
{Q^2 \over 2 M^2 } \ \beta_1 (Q^2) \ {1 \over x}  +
{2 \over x}
\sum_{n=0,2,4,..} \biggl( {1 \over x^n} \biggr) \int_0^1 dy \ y^n g_1 (y, Q^2) 
\nonumber \\
\alpha_2 (x, Q^2)
&=& 
{Q^2 \over 2 M^2 } \ \beta_2 (Q^2) \ {1 \over x} +
{2 \over x^3}
\sum_{n=0,2,4,..} \biggl( {1 \over x^n} \biggr) \int_0^1 dy \ y^{n+2} 
g_2 (y, Q^2) 
\nonumber \\
\label{eqcm}
\end{eqnarray}
with $x={1 \over \omega}$.

These equations form the basis for the spin sum rules for polarized photon 
nucleon scattering.
We next outline the derivation of the Bjorken \cite{bj} and 
Ellis-Jaffe \cite{ej}
sum rules for isovector and flavour-singlet parts of $g_1$ 
in polarized deep inelastic scattering,
the Burkhardt-Cottingham sum rule for $G_2$ \cite{cottingham},
and 
the Drell-Hearn-Gerasimov sum rule for polarized photoproduction \cite{dhg}.
Each of these spin sum rules assumes no subtraction at infinity.

\subsection{Deep inelastic spin sum rules}

Sum rules for polarized deep inelastic scattering are derived by combining
the dispersion relation expressions (\ref{eqcm}) 
with the light cone operator production expansion.
When $Q^2 \rightarrow \infty$ the leading contribution to the spin dependent 
part of the forward
Compton amplitude comes from the nucleon matrix elements of a tower of gauge 
invariant local operators multiplied by Wilson coefficients
\footnote
{Note that, 
 for simplicity, in this discussion we consider the case of a single quark 
 flavour with unit charge.  
 The results quoted in Section 3.2 below
 include the extra steps of using the full electromagnetic current of QCD.}
:
\begin{equation}
T_{\mu \nu}^A =
i \epsilon_{\mu \nu \lambda \sigma} q^{\lambda} \sum_{n=0,2,4,..}
\biggl( - {2 \over q^2} \biggr)^{n+1}
q^{\mu_1} q^{\mu_2} ... q^{\mu_n} 
\sum_{i=q,g}
\Theta^{(i)}_{\sigma \{ \mu_1 ... \mu_n \} }
E^i_n ({Q^2 \over \mu^2}, \alpha_s)
\label{eqcca}
\end{equation}
where
\begin{equation}
\Theta^{(q)}_{\sigma \{\mu_1 ... \mu_n \} } \equiv i^n 
\bar{\psi} \gamma_{\sigma} \gamma_5 D_{ \{ \mu_1} ... D_{\mu_n \} } \psi
- {\rm traces}  
\label{eqccb}
\end{equation}
and
\begin{equation}
\Theta^{(g)}_{\sigma \{\mu_1 ... \mu_n \} } \equiv 
i^{n-1} 
\epsilon_{\alpha \beta \gamma \sigma}
G^{\beta \gamma}
 D_{ \{ \mu_1} ... D_{\mu_{n-1}} G_{\ \mu_{n} \}}^{\alpha}
- {\rm traces}  
\label{eqccc}
\end{equation}
Here $D_{\mu}$ is the gauge covariant derivative and the sum over even
values of $n$ in 
Eq.(\ref{eqcca})
reflects the crossing symmetry properties of $T_{\mu \nu}$.
The functions 
$E^q_n ({Q^2 \over \mu^2}, \alpha_s)$ and
$E^g_n ({Q^2 \over \mu^2}, \alpha_s)$ are the respective Wilson coeffients.
The operators in Eq.(\ref{eqcca}) 
may each be written as the sum of a 
totally symmetric operator and an operator with mixed symmetry
\begin{equation}
\Theta_{\sigma \{\mu_1 ... \mu_n \} }
=
\Theta_{ \{ \sigma \mu_1 ... \mu_n \} }
+
\Theta_{ [ \sigma , \{ \mu_1 ] ... \mu_n \} }
\label{eqccd}
\end{equation}
These operators have the matrix elements:
\begin{eqnarray}
\langle p, s | \Theta_{\{\sigma \mu_1 ... \mu_n \}} | p,s \rangle
&=&
\{ s_{\sigma} p_{\mu_1} ... p_{\mu_n} 
 + s_{\mu_1} p_{\sigma} p_{\mu_2} ... p_{\mu_n} 
 + ... \}
{a_n \over n+1}
\nonumber \\
\langle p,s | \Theta_{[ \{ \sigma \mu_1 ]... \mu_n \}} | p,s \rangle 
&=&
\{ ( s_{\sigma} p_{\mu_1} - s_{\mu_1} p_{\sigma} ) p_{\mu_2} ... p_{\mu_n} 
\nonumber \\
& & 
\ \ \ \
 + (s_\sigma p_{\mu_2} - s_{\mu_2} p_{\sigma}) p_{\mu_1} ... p_{\mu_n} 
 + ... \}
{d_n \over n+1}
\label{eqcce}
\end{eqnarray}
Now define
${\tilde a}_n = a_n^{(q)} E_{1 n}^q +  a_n^{(g)} E_{1 n}^g$ 
and
${\tilde d}_n = d_n^{(q)} E_{2 n}^q +  d_n^{(g)} E_{2 n}^g$
where $E_{1 n}^i$ and $E_{2 n}^i$ are 
the Wilson coefficients 
for $a_n^i$ and $d_n^i$ respectively.
Combining equations (\ref{eqcca}) and (\ref{eqcce}) one obtains equations 
for $\alpha_1$ and $\alpha_2$:
\begin{eqnarray}
\alpha_1(x,Q^2) + \alpha_2(x,Q^2) 
&=&
\sum_{n=0,2,4,...} {{\tilde a}_n + n {\tilde d}_n \over n+1} {1 \over x^{n+1}}
\nonumber \\
\alpha_2(x,Q^2) 
&=&
\sum_{n=2,4,...} {n ({\tilde d}_n - {\tilde a}_n) \over n+1} {1 \over x^{n+1}}
\label{eqccf}
\end{eqnarray}
These equations are compared with the Taylor series expansions
(\ref{eqcm}), 
whence we obtain
the moment sum rules for $g_1$ and $g_2$:
\begin{equation}
\int_0^1 dx x^n g_1 = {1 \over 2} {\tilde a}_n
\end{equation}
for $= 0,2,4,...$
and
\begin{equation}
\int_0^1 dx x^n g_2 = {1 \over 2} {n \over n+1} ({\tilde d}_n - {\tilde a}_n)
\label{eqccg}
\end{equation}
for $n = 2,4,6,...$
\\
Note:
\begin{enumerate}
\item
The first moment of $g_1$ is given by the nucleon matrix element of 
the axial vector current ${\bar \psi} \gamma_{\sigma} \gamma_5 \psi$.
There is no twist-two, spin-one, gauge-invariant, local gluon operator 
to contribute to the first moment of $g_1$ \cite{jaffem}.
\item
The potential subtraction term ${Q^2 \over 2 M} \beta_1 (Q^2)$ 
in the dispersion relation (\ref{eqcl})
multiplies a ${1 \over x}$ term in the series expansion
on the left hand side, 
and thus provides a 
potential correction factor to sum rules for the first moment of $g_1$.
It follows that the first moment of $g_1$ measured in polarized
deep inelastic scattering measures the nucleon matrix element of
the axial vector current
up to this potential ``subtraction at infinity'' term, 
which corresponds to the residue of any $J=1$ 
fixed pole with nonpolynomial residue contribution to the real part of $A_1$.
\item
There is no ${1 \over x}$ term in the operator product expansion formula 
(\ref{eqccf}) for $\alpha_2 (x, Q^2)$.
This is matched by the lack of any ${1 \over x}$ term in the unsubtracted
version of the dispersion relation (\ref{eqcm}). 
The operator product expansion provides no information about the first 
moment of $g_2$ without 
additional assumptions concerning analytic continuation and the $x \sim 0$ 
behaviour of $g_2$ \cite{jaffeg2}
--- see the discussion about the Burkhardt Cottingham sum rule in Section 3.3.
\end{enumerate}
If there are finite subtraction constant corrections 
to one (or more) spin sum rules, 
one can include the correction by re-interpreting 
the relevant structure function as a distribution 
with the subtraction constant included as the coefficient of a 
$\delta (x)$ term \cite{bgj}.

\subsection{$g_1$ spin sum rules in polarized deep inelastic scattering}

The value of $g_A^{(0)}$ extracted from polarized deep inelastic 
scattering is obtained as follows.
One includes the sum over quark charges squared in $W_{\mu \nu}$ 
and assumes no twist-two subtraction constant 
($\beta_1 (Q^2) = O(1/Q^4)$).
The first moment of  the structure function $g_1$
is then related 
to the scale-invariant axial charges of the target nucleon by
\begin{eqnarray}
\int_0^1 dx \ g_1^p (x,Q^2) &=&
\Biggl( {1 \over 12} g_A^{(3)} + {1 \over 36} g_A^{(8)} \Biggr)
\Bigl\{1 + \sum_{\ell\geq 1} c_{{\rm NS} \ell\,}
\alpha_s^{\ell}(Q)\Bigr\} \nonumber \\
&+& {1 \over 9} g_A^{(0)}|_{\rm inv}
\Bigl\{1 + \sum_{\ell\geq 1} c_{{\rm S} \ell\,}
\alpha_s^{\ell}(Q)\Bigr\}
\ + \ {\cal O}({1 \over Q^2}).
\label{eqcch}
\end{eqnarray}
Here $g_A^{(3)}$, $g_A^{(8)}$ and $g_A^{(0)}|_{\rm inv}$ are
the isovector, SU(3)
octet and scale-invariant  flavour-singlet axial charges respectively.
The flavour non-singlet $c_{{\rm NS} \ell}$ and singlet
$c_{{\rm S} \ell}$
Wilson coefficients are calculable in $\ell$-loop perturbative QCD
\cite{larin}.

Note that the first moment of $g_1$ is constrained by low energy weak
interactions.
For proton states $|p,s\rangle$ with momentum $p_\mu$ and spin $s_\mu$
\begin{eqnarray}
2 m s_{\mu} \ g_A^{(3)} &=&
\langle p,s |
\left(\bar{u}\gamma_\mu\gamma_5u - \bar{d}\gamma_\mu\gamma_5d \right)
| p,s \rangle   \nonumber \\
2 m s_{\mu} \ g_A^{(8)} &=&
\langle p,s |
\left(\bar{u}\gamma_\mu\gamma_5u + \bar{d}\gamma_\mu\gamma_5d
                   - 2 \bar{s}\gamma_\mu\gamma_5s\right)
| p,s \rangle
\label{eqcci}
\end{eqnarray}
Here $g_A^{\scriptscriptstyle (3)} = 1.267 \pm 0.004$
is the isovector axial charge measured in neutron beta-decay;
$g_A^{\scriptscriptstyle (8)} = 0.58 \pm 0.03$
is the octet charge measured independently in 
hyperon beta decays (using SU(3)) \cite{su3}.
(The assumption of good SU(3) here
 is supported 
 by the recent KTeV measurement \cite{ktev} of 
 the $\Xi^0$ beta decay $\Xi^0 \rightarrow \Sigma^+ e {\bar \nu}$.)

The scale-invariant flavour-singlet axial charge 
$g_A^{(0)}|_{\rm inv}$ 
is defined by 
\begin{equation}
2m s_\mu g_A^{(0)}|_{\rm inv} = 
\langle p, s|
\ E(\alpha_s) J^{GI}_{\mu5} \ |p, s \rangle 
\label{eqccj}
\end{equation}
where 
\begin{equation}
J^{GI}_{\mu5} = \left(\bar{u}\gamma_\mu\gamma_5u
                  + \bar{d}\gamma_\mu\gamma_5d
                  + \bar{s}\gamma_\mu\gamma_5s\right)_{GI}
\label{eqcck}
\end{equation} 
is the
gauge-invariantly renormalized singlet axial-vector operator 
and
\begin{equation}
E(\alpha_s) = \exp \int^{\alpha_s}_0 \! d{\tilde \alpha_s}\, 
\gamma({\tilde \alpha_s})/\beta({\tilde \alpha_s})
\label{eqccl}
\end{equation}
is a renormalization group factor 
which corrects 
for the (two loop) non-zero anomalous dimension 
$\gamma(\alpha_s)$ ($= f {\alpha_s^2 \over \pi^2} + {\cal O}(\alpha_s^3)$) 
of $J_{\mu 5}^{GI}$
\cite{kod}.
Here $\beta (\alpha_s)$ is the QCD beta function. We are free to choose 
the QCD coupling $\alpha_s(\mu)$ at either a hard or a soft scale $\mu$.
The singlet axial charge 
$g_A^{(0)}|_{\rm inv}$ 
is independent of the renormalization scale 
$\mu$ and corresponds 
to $g_A^{(0)}(Q^2)$ evaluated in the limit $Q^2 \rightarrow \infty$. 
If we take $\alpha_s (\mu_0^2) \sim 0.6$ as typical of the infrared
region of QCD, then the renormalization group factor
$E(\alpha_s) \simeq 1 - 0.13 - 0.03 = 0.84$ 
where -0.13 and -0.03 
are the ${\cal O}(\alpha_s)$ and ${\cal O}(\alpha_s^2)$ corrections 
respectively.

In the isovector channel the Bjorken sum rule \cite{bj,larin}
\begin{equation}
I_{Bj} = 
\int_0^1 dx \Biggl( g_1^p - g_1^n \Biggr)
=
\frac{g_A^{(3)}}{6} 
\left[1 - \frac{\alpha_s}{\pi} - 3.58 \left(\frac{\alpha_s}{\pi} \right)^2 
        - 20.21 \left(\frac{\alpha_s}{\pi} \right)^3 \right] 
\label{eqccm}
\end{equation}
has been confirmed at the level of 10\%.
Using the value $g_A^{(8)} = 0.58 \pm 0.03$ 
from hyperon beta-decays 
(and assuming no subtraction constant correction)
the polarized deep inelastic data implies
\begin{equation}
\left. g^{(0)}_A \right|_{\rm pDIS} = 0.2 - 0.35
\end{equation}
for the flavour singlet (Ellis Jaffe) moment
corresponding to the
polarized strangeness
$\Delta s = -0.10 \pm 0.04$ quoted in Section 1.

The small $x$ extrapolation of $g_1$ data is presently the largest source 
of experimental error on measurements of the nucleon's axial charges from 
deep inelastic scattering.
We refer to Ziaja \cite{ziaja} 
for a recent discussion of perturbative QCD predictions for the small $x$ 
behaviour of $g_1$ in deep inelastic scattering.

Note that polarized deep inelastic scattering experiments measure $g_1$ 
between some small but finite value $x_{\rm min}$ 
and an upper value 
$x_{\rm max}$ which is close to one.
Deep inelastic measurements of $g_A^{(3)}$ and $g_A^{(0)}$ 
involve a smooth extrapolation of the $g_1$ data to $x=0$ 
which is motivated either by Regge theory or by perturbative QCD.
As we decrease $x_{\rm min} \rightarrow 0$ we measure the first moment
\begin{equation}
\Gamma \equiv \lim_{x_{\rm min} \rightarrow 0} \ 
\int^1_{x_{\rm min}} dx \ g_1 (x,Q^2).
\label{eqccn}
\end{equation}
Polarized deep inelastic experiments cannot, even in principle, measure at 
$x=0$ with finite $Q^2$.
They miss any possible $\delta (x)$ terms which might exist in $g_1$ at 
large $Q^2$. 
That is, they miss any potential (leading twist) 
fixed pole corrections
and/or zero mode (topological) contributions to $g_A^{(0)}|_{\rm inv}$.

\subsection{The Burkhardt Cottingham sum rule}

The Burkhardt Cottingham sum rule \cite{cottingham} reads:
\begin{equation}
\int_{Q^2/2M}^{\infty} d \nu G_2 (Q^2, \nu) 
= {2 M^3 \over Q^2} \int_0^1 dx g_2 
= 0
\label{eqcco}
\end{equation}
For deep inelastic scattering, this sum rule is derived
by assuming that the moment formula (\ref{eqccg})
can be analytically continued to $n=0$.
In general, the Burkhardt Cottingham sum rule is derived 
by assuming no $\alpha \geq 0$
singularity in $G_2$
(or, equivalently, no ${1 \over x}$ or more singular small 
 behaviour in $g_2$)
and no ``subtraction at infinity'' 
(from an
 $\alpha = J = 0$ fixed pole in the real part of $G_2$)
\cite{jaffeg2}.
The most precise measurements of $g_2$ 
to date
in polarized deep inelastic scattering 
come from
the SLAC E-155 and E-143 experiments, which report
$\int_{0.02}^{0.8} dx  \ g_2^p = -0.042 \pm 0.008$ 
for the proton and
$\int_{0.02}^{0.8} dx  \ g_2^d = -0.006 \pm 0.011$ 
for the deuteron at $Q^2=5$GeV$^2$ \cite{slacg2}.
New, even more accurate, 
measurements 
of $g_2$
(for the neutron using a $^3$He target)
are becoming available at 
Jefferson Laboratory \cite{jlabbc} for $Q^2$ between 0.1 and 0.9 GeV$^2$.
Further measurements to test the Burkhardt-Cottingham sum rule would 
be most valuable, particularly given the SLAC proton result quoted above.

\subsection{The Drell Hearn Gerasimov sum rule}

The Drell-Hearn-Gerasimov sum-rule \cite{dhg} for spin dependent 
photoproduction
relates the difference of the two cross-sections for the absorption
of a real photon with spin anti-parallel $\sigma_{1 \over 2}$ and parallel 
$\sigma_{3 \over 2}$ to the target spin to the square of the anomalous 
magnetic 
moment of the target.
It is derived by setting $\nu=0$ in the dispersion relation for $A_1$, 
Eq.(\ref{eqcg}).
For small photon energy $\nu \rightarrow 0$
\begin{equation}
A_1(0,\nu) = - {1 \over 4} \kappa^2 + O(\nu^2) ,
\label{eqccp}
\end{equation}
where $\kappa$ is the anomalous magnetic moment of the target.
This low-energy theorem follows from Lorentz invariance and 
electromagnetic gauge invariance
(plus the existence of a finite mass gap between the 
 ground state and continuum
 contributions to forward Compton scattering) \cite{brod69,low}.
The Drell-Hearn-Gerasimov sum rule reads:
\begin{equation}
\int_{\rm threshold}^{\infty} {d \nu \over \nu}
( \sigma_{1 \over 2} - \sigma_{3 \over 2} )
= {8 \pi^2 \alpha \over M^2} 
\int_{\rm threshold}^{\infty} {d \nu \over \nu}
G_1
= - {2 \pi^2 \alpha \over M^2} \kappa^2
\label{eqccq}
\end{equation}
The sum rule 
follows from the very general principles of causality, unitarity, 
Lorentz and electromagnetic gauge invariance and one assumption: 
that the $g_1$ spin structure function satisfies an unsubtracted 
dispersion relation.
Modulo the no-subtraction hypothesis, 
the Drell-Hearn-Gerasimov sum-rule is valid for a target of 
arbitrary spin $S$, whether elementary or composite \cite{brod69}
-- for a review see \cite{bassdhg}.

The integral in Eq.(\ref{eqccq}) converges for each of the leading 
Regge contributions 
(discussed below  Eq.(\ref{eqbl})).
If the sum rule were observed 
to fail 
(with finite integral)
the interpretation would be a ``subtraction at infinity'' 
induced by a $J=1$ fixed pole in the real part of the spin amplitude
$A_1$ 
\cite{abarbanel}.

Experimental investigations of the Drell-Hearn-Gerasimov sum rule are
being carried out at several laboratories: 
ELSA and MAMI, JLab, GRAAL, LEGS@BNL, and SPRING.
Preliminary results \cite{helbing} from the ELSA-MAMI experiments 
suggest that the contribution to the DHG integral 
for a proton target
from energies $\nu < 3$GeV 
exceeds
the total sum rule prediction (-204.5$\mu$b) by about 5-10\%.
Phenomenological estimates 
suggest that about $+25 \pm 10 \mu$b of the sum rule may reside at 
higher energies
\cite{brisudova}.
However it should be noted that any 10\% fixed pole correction would 
be competitive with this high energy contribution within the errors.
Further measurements, including at higher energy, would be valuable.
These measurements could be carried out at SLAC 
or using a 
future polarized $ep$ collider \cite{bassdreic}.
In addition to mapping out spin dependent Regge theory and placing an 
upper bound on the the high energy contribution 
to the Drell-Hearn-Gerasimov sum rule high energy measurements of $G_1$ 
in polarized photoproduction would provide a baseline for investigations
of perturbative QCD motivated small $x$ behaviour in $g_1$.
The transition region between polarized photoproduction and deep inelastic
$Q^2$ is expected to reveal much larger changes in the effective intercept
for small $x$ physics than those observed in the unpolarized structure function
$F_2$ \cite{bassdreic}.

\section{Fixed Poles}

Fixed poles are exchanges in Regge phenomenology with no $t$ dependence:
the trajectories are described by
$J = \alpha(t) = 0$ or 1 for all $t$
\cite{fixedpole}.
For example, for fixed $Q^2$
a $t-$independent real constant term in the spin amplitude $A_1$ would 
correspond to a $J=1$ fixed pole.
Fixed poles are excluded in hadron-hadron scattering 
by unitarity but are not
excluded from Compton amplitudes (or parton distribution functions)
because these are calculated only to lowest order in the current-hadron 
coupling.
Indeed, there are two famous examples where fixed poles are required: 
(by current algebra) in the Adler sum rule for W-boson nucleon scattering,
and to reproduce the Schwinger term sum rule for the longitudinal structure 
function
measured in unpolarized deep inelastic $ep$ scattering.
We review the derivation of 
these fixed pole contributions, 
and then discuss potential fixed pole 
corrections 
to the Burkhardt-Cottingham, $g_1$ and Drell-Hearn-Gerasimov sum-rules.
\footnote
{We refer to \cite{efremov} for a 
 recent discussion of an ``$x=0$'' fixed pole contribution 
 to the twist 3, chiral-odd
 structure function $e(x)$.}
Fixed poles in the real part of the forward Compton amplitude have the 
potential to induce  
``subtraction at infinity'' 
corrections to sum rules for photon nucleon (or lepton nucleon) scattering.
For example, a $\nu$ independent term in the real part of $A_1$ would 
induce a subtraction constant correction to the spin sum rule for the first 
moment of $g_1$.
Bjorken scaling at large $Q^2$ constrains the $Q^2$ dependence of 
the residue 
of any fixed pole in the real of the forward Compton amplitide
(e.g. $\beta_1(Q^2)$ and $\beta_2(Q^2)$ in the dispersion relations 
 (\ref{eqcl}) ).
To be consistent with scaling
these residues must decay as or faster 
than $1/Q^2$ as $Q^2\to\infty$.  
That is, they must be nonpolynomial in $Q^2$.

\subsection{Adler sum rule}

The first example we consider is the Adler sum rule for W-boson nucleon
scattering \cite{adler}:
\begin{eqnarray}
\int_{Q^2/2M}^{+ \infty} d \nu 
\biggl[ 
W_2^{\bar{\nu} p} (\nu, Q^2) - W_2^{\nu p} (\nu, Q^2)
\biggr]
&=&
\int_0^1 {dx \over x}
\biggl[ 
F_2^{\bar{\nu} p} (x, Q^2) - F_2^{\nu p} (x, Q^2)
\biggr]
\nonumber \\
&=&
\begin{array}{c}
4 - 2 \cos^2 \theta_c \ \ \ \ \ ({\rm BCT}) \\
2 \ \ \ \ \ \ \ \ \ \ \ \ \ \ \ \ \ \ \ ({\rm ACT}) 
\end{array} 
\label{eqda}
\end{eqnarray}
Here
$\theta_c$ is the Cabibbo angle, and
BCT and ACT refer to below and above the charm production threshold.

The Adler sum rule is derived from current algebra.
The right hand side of the sum rule is the coefficient of a $J=1$ 
fixed pole term
\begin{equation}
{i \over \pi} f_{abc} \ F_c \
\biggl[ (p_{\mu} q_{\nu} + q_{\mu} p_{\nu}) - M \nu g_{\mu \nu} 
\biggr] / Q^2
\end{equation}
in the imaginary part of the forward Compton amplitude for W-boson nucleon 
scattering \cite{heimann}.
This fixed pole term is required by the 
commutation relations between the charge raising and lowering weak currents
\begin{eqnarray}
q_{\mu} T^{\mu \nu}_{ab} 
&=& 
- {1 \over \pi}
\int d^4 x \  e^{i q.x} \ 
\langle p,s | \biggl[ J_a^{0} (x), 
                     J^{\nu}_b (0) \biggr] | p,s \rangle 
\delta (x^0)
\nonumber \\
&=& - {i \over \pi} f_{abc} \langle ps |  J_c^{\nu} (0)| ps \rangle 
\end{eqnarray}
Here
$F_c$ is a generalized form factor at zero momentum transfer:
\begin{equation}
\langle p,s | J_c^{\nu} (0) | p,s \rangle \equiv p^{\nu} F_c
\end{equation}
The fixed pole term appears in lowest order perturbation theory, 
and is not
renormalized because it is a consequence of current algebra.
The Adler sum rule is protected against radiative QCD corrections.

\subsection{Schwinger term sum rule}

Our second example is the Schwinger term sum rule \cite{bgj}
which relates the 
logarithmic integral in $\omega$ 
(or Bjorken $x$) 
of the longitudinal structure function $F_L (\omega, Q^2)$
($F_L = {1 \over 2} \omega F_2 - F_1$)
measured in unpolarized deep inelastic scattering to the 
target matrix element of the operator Schwinger term ${\cal S}$
defined through the equal-time
commutator of the electromagnetic charge and current densities
\begin{equation}
\langle p,s | 
\biggl[ J_0({\vec y}, 0), J_i(0) \biggr] 
| p,s \rangle
= i \ \partial_i \ \delta^3 ({\vec y}) \ {\cal S} .
\label{eqdc} 
\end{equation}
The Schwinger term sum rule reads 
\begin{equation}
{\cal S}
=
\lim_{Q^2 \rightarrow \infty}
\biggl[ 
4 \int_{1}^{\infty} 
  {d \omega \over \omega} {\tilde F}_L (\omega, Q^2) 
-
4 \sum_{\alpha >0} \gamma (\alpha,Q^2) / \alpha
- C(q^2) 
\biggr] 
\label{eqdd}
\end{equation}
Here $C(Q^2)$ is the nonpolynomial residue of any $J=0$ fixed 
pole contribution in the real part of $T_2$
and
\begin{equation}
{\tilde F}_L (\omega, Q^2) 
= 
F_L (\omega, Q^2)
-
\sum_{\alpha  \geq 0} \gamma (\alpha, Q^2) \omega^{\alpha} .
\label{eqde}
\end{equation}
The integral in Eq.(\ref{eqdd}) 
is convergent because ${\tilde F}_L (\omega, Q^2)$ 
is defined with all Regge contributions with effective intecept 
greater than or equal to zero removed from $F_L (Q^2, \omega)$.
The Schwinger term ${\cal S}$ vanishes in vector gauge theories 
like QCD.
Since $F_L (\omega, Q^2)$ is positive definite, it follows that
QCD possesses the required non-vanishing $J=0$ fixed pole in the
real part of $T_2$.

\subsection{Burkhardt Cottingham sum rule}

The third example, and the first in connection with spin, is the Burkhardt 
Cottingham sum rule for the first moment of $g_2$
\cite{cottingham}:
\begin{equation}
\int_{Q^2/2M}^{\infty} d \nu \ G_2 (Q^2, \nu) 
= {2 M^3 \over Q^2} \int_0^1 dx g_2 
= 0
\nonumber
\end{equation}
Suppose that future experiments find that the sum rule is violated and 
that the integral is finite.
The conclusion \cite{jaffeg2}
would be a $J=0$ fixed pole with nonpolynomial residue in the real part 
of $A_2$. 
To see this 
work at fixed $Q^2$ and assume that all Regge-like singularities 
contributing to $A_2 (\nu, Q^2)$ have intercept less than zero so that
\begin{equation}
A_2 (\nu, Q^2) \sim \nu^{-1 - \epsilon}
\label{eqdf}
\end{equation}
as $\nu \rightarrow \infty$ for some $\epsilon < 0$.
Then the large $\nu$ behaviour of $A_2$ is obtained
by taking 
$\nu \rightarrow \infty$ under the $\nu'$ integral giving
\begin{equation}
A_2(Q^2, \nu) \sim - {2 \over \pi \nu} 
\int_{Q^2/2M}^{\infty} d \nu' \ {\rm Im} A_2 (Q^2, \nu')
\label{eqdg}
\end{equation}
which contradicts the assumed behaviour unless the integral vanishes;
hence the sum rule.
{\it If} there is an $\alpha(0)=0$ fixed pole in 
the real part of $A_2$ 
the fixed pole will not 
contribute to ${\rm Im} A_2$ 
and therefore not spoil the convergence of the integral. 
One finds
\begin{equation}
\beta_2(Q^2) \sim - {2 \over \pi M} 
\int_{Q^2/2M}^{\infty} d \nu' \ {\rm Im} A_2 (Q^2, \nu')
\label{eqdh}
\end{equation}
for the residue of any $J=0$ fixed pole coupling to $A_2(Q^2, \nu)$.

\subsection{$g_1$ spin sum rules}

Scaling requires that any fixed pole correction to the Ellis Jaffe $g_1$ 
sum rule must have nonpolynomial residue.
Through Eq.(\ref{eqcm}), the fixed pole coefficient $\beta_1 (Q^2)$
must decay as or faster than $O(1/Q^2)$ as $Q^2 \rightarrow \infty$.
The coefficient is further constrained 
by the requirement that $G_1$ contains no kinematic singularities
(for example at $Q^2=0$).
In Section 5 we will identify a potential leading-twist topological
$x=0$ contribution to the first moment of $g_1$ through analysis of 
the axial 
anomaly contribution to $g_A^{(0)}$.
This zero-mode topological contribution (if finite) generates a
leading twist fixed pole correction to the flavour-singlet part of
$\int_0^1 dx g_1$.
{\it If} present, this fixed pole will also violate 
the Drell-Hearn-Gerasimov sum rule 
(since the two sum rules are derived from $A_1$)
unless the underlying dynamics suppress the fixed pole's residue at $Q^2=0$.

At this point
it is interesting to consider the $g_1$ spin structure function of a
polarized real photon.
(Assuming no fixed pole correction)
the first moment of $g_1^{\gamma}$ of a real photon vanishes 
\begin{equation}
\int_0^1 \ dx  \ g_1^{\gamma} (x, Q^2) = 0
\label{eqdia}
\end{equation}
independent of the virtuality $Q^2$ of the photon that it is probed with
\cite{bassbs,g1gamma}.
This result is non-perturbative.
There are two derivations.
In the first we treat the real photon as the beam and the virtual photon,
and apply the Drell-Hearn-Gerasimov sum rule. 
The anomalous magnetic moment of a photon vanishes to all orders because of 
Furry's theorem.
Alternatively (for large $Q^2$),
we can treat the deeply virtual photon as the beam and apply the operator 
product expansion.
The sum rule (\ref{eqdia}) holds to all orders in perturbation theory and 
at every twist.
If there is a fixed pole correction to the polarized real photon 
spin sum rule (\ref{eqdia})
then the correction will affect 
both the deep inelastic first moment (applied to the deeply virtual photon)
and
Drell-Hearn-Gerasimov (applied to the real photon) sum rules 
for the polarized photon system.
Measurements of $g_1^{\gamma}$ might be possible with a polarized 
$e \gamma$ collider \cite{nlc}.

Note that any fixed pole correction to the Drell-Hearn-Gerasimov sum rule 
is most probably a non-perturbative effect.
The sum rule (\ref{eqccq})
has been verified 
to $O(\alpha^2)$ for all 
$2 \rightarrow 2$ processes $\gamma a \rightarrow bc$
where $a$ is either a real lepton, quark, gluon 
or elementary Higgs target \cite{acm},
and for electrons in QED to $O(\alpha^3)$ \cite{dicus}.

One could test for a fixed pole correction to the Ellis-Jaffe moment 
through a precision measurement of the flavour singlet axial charge 
from an independent process where one is not sensitive to theoretical 
assumptions about the presence or absence of a $J=1$ fixed pole in $A_1$.
Here the natural choice is elastic neutrino proton scattering 
\cite{tayloe,garvey}
where the parity violating part of the cross-section includes a direct
weak interaction measurement of the scale invariant flavour-singlet 
axial charge $g_A^{(0)}|_{\rm inv}$, 
or through parity violation in light atoms \cite{fortson,ellis}.

The subtraction constant fixed pole correction hypothesis could 
also, in principle, be tested through measurement of the real 
part of the spin dependent part of the forward deeply virtual 
Compton amplitude. 
While this measurement may seem extremely difficult at 
the present time one should not forget that Bjorken believed 
when writing his original 
Bjorken sum rule paper that the sum rule would never be tested \cite{bj}!

\subsection{$\nu p$ elastic scattering}

Neutrino proton elastic scattering measures the proton's weak axial 
charge $\gaz$ through elastic Z$^0$ exchange.
Because of anomaly cancellation in the Standard Model
the weak neutral current couples to the combination $u-d+c-s+t-b$,
{\it viz.}
\begin{equation}
J_{\mu5}^Z\
=\ \smallfrac{1}{2} \biggl\{\,\sum_{q=u,c,t} - \sum_{q=d,s,b}\,\biggr\}\:
        \bar{q}\gamma_\mu\gamma_5q
\label{eqdj}
\end{equation}
It measures the combination
\begin{equation}
2\gaz = \bigl( \Delta u - \Delta d - \Delta s \bigr)
       + \bigl( \Delta c - \Delta b + \Delta t \bigr)
\label{eqdk}
\end{equation}
where $\Delta q$ refers to the expectation value
\[ \langle p,s|\,\bar{q}\gamma_\mu\gamma_5q\,|p,s \rangle
  = 2 M s_\mu\Delta q \]
for a proton of spin $s_\mu$ and mass $M$.
Heavy quark renormalization group arguments 
can be used 
to calculate the heavy $t$, $b$ and $c$ quark contributions to $\gaz$.
The full NLO result is \cite{bcsta}
\begin{equation}
2\gaz\, =\, \bigl(\Delta u - \Delta d - \Delta s\bigr)_{\rm inv}
           +\hsp{0.2} {\cal H}\hsp{0.1}\bigl(
               \Delta u + \Delta d + \Delta s\bigr)_{\rm inv}
    +\, O(m_{t,b,c}^{-1})
\label{eqdka}
\end{equation}
where ${\cal H}$ is a polynomial in the running couplings
$\run{h}$,
\begin{eqnarray}
{\cal H}\, =\,& &\smallfrac{6}{23\pi}\bigl(\run{b}-\run{t}\bigr)
             \Bigl\{1 + \smallfrac{125663}{82800\pi}\run{b}
                      + \smallfrac{6167}{3312\pi}\run{t}
                      - \smallfrac{22}{75\pi}\run{c}  \Bigr\}
\nonumber \\
& &\phantom{+\ \Biggl[} - \smallfrac{6}{27\pi} \run{c}
                      - \smallfrac{181}{648 \pi^2}\run{c}^2
                      + O\bigl(\run{t,b,c}^3\bigr)
\label{eqdl}
\end{eqnarray}
Here $(\Delta q)_{\rm inv}$ denotes the scale-invariant version of 
$\Delta q$ and $\run{h}$ 
denotes Witten's renormalization-group-invariant running couplings
for heavy quark physics \cite{witten,bcstb}.
Taking $\widetilde{\alpha}_t = 0.1$, $\widetilde{\alpha}_b = 0.2$ 
and $\widetilde{\alpha}_c = 0.35$ in (\ref{eqdl}), one finds a small 
heavy-quark correction factor ${\cal H}= -0.02$, with LO terms dominant.

Modulo the small heavy-quark corrections quoted above, a precision 
measurement of $g_A^{(Z)}$,
together with
$g_A^{(3)}$ and $g_A^{(8)}$,
would
provide a weak interaction determination of $(\Delta s)_{\rm inv}$,
complementary to the deep inelastic measurement (\ref{eqa2}).
The $\nu p$ elastic  measurement may be possible \cite{tayloe} at
FNAL using the mini-BooNE set-up with small duty factor 
($\sim 10^{-5}$)
neutrino beam to control backgrounds.
The estimated error on the strange quark polarization one could 
extract from this experiment is $ \sim 0.03$, 
competitive with the error from present polarized deep inelastic measurements.

\section{The axial anomaly, gluon topology and $g_A^{(0)}$}

We next discuss the role of the axial anomaly in the interpretation of
$g_A^{(0)}$.

\subsection{The axial anomaly}

In QCD one has to consider the effects of renormalization.
The flavour singlet axial vector current $J_{\mu 5}^{GI}$ 
in Eq.(\ref{eqcck})
satisfies the anomalous divergence equation 
\cite{anomaly,rjc}
\begin{equation}
\partial^\mu J^{GI}_{\mu5}
= 2f\partial^\mu K_\mu + \sum_{i=1}^{f} 2im_i \bar{q}_i\gamma_5 q_i
\label{e1}
\end{equation}
where
\begin{equation}
K_{\mu} = {g^2 \over 32 \pi^2}
\epsilon_{\mu \nu \rho \sigma}
\biggl[ A^{\nu}_a \biggl( \partial^{\rho} A^{\sigma}_a 
- {1 \over 3} g 
f_{abc} A^{\rho}_b A^{\sigma}_c \biggr) \biggr]
\label{e2}
\end{equation}
is a renormalized version of the gluonic Chern-Simons
current
and the number of light flavours $f$ is $3$.
Eq.(\ref{e1}) 
allows us to write
\begin{equation}
J_{\mu 5}^{GI} = J_{\mu 5}^{\rm con} + 2f K_{\mu}
\label{e3}
\end{equation}
where $J_{\mu 5}^{\rm con}$ and $K_{\mu}$ satisfy the
divergence equations
\begin{equation}
\partial^\mu J^{\rm con}_{\mu5}
= \sum_{i=1}^{f} 2im_i \bar{q}_i\gamma_5 q_i
\label{e4}
\end{equation}
and
\begin{equation}
\partial^{\mu} K_{\mu} 
= {g^2 \over 32 \pi^2} G_{\mu \nu} {\tilde G}^{\mu \nu}.
\label{e5}
\end{equation}
Here
${g^2 \over 32 \pi^2} G_{\mu \nu} {\tilde G}^{\mu \nu}$
is the topological charge density.
The partially conserved current is scale invariant 
and 
the scale dependence of $J_{\mu 5}^{GI}$ is carried entirely
by $K_{\mu}$.
When we make a gauge transformation $U$ 
the gluon field transforms as
\begin{equation}
A_{\mu} \rightarrow U A_{\mu} U^{-1} + {i \over g} (\partial_{\mu} U) U^{-1}
\label{e6}
\end{equation}
and the operator $K_{\mu}$
transforms as
\begin{equation}
K_{\mu} \rightarrow K_{\mu} 
+ i {g \over 8 \pi^2} \epsilon_{\mu \nu \alpha \beta}
\partial^{\nu} 
\biggl( U^{\dagger} \partial^{\alpha} U A^{\beta} \biggr)
+ {1 \over 24 \pi^2} \epsilon_{\mu \nu \alpha \beta}
\biggl[ 
(U^{\dagger} \partial^{\nu} U) 
(U^{\dagger} \partial^{\alpha} U)
(U^{\dagger} \partial^{\beta} U) 
\biggr].
\label{e7}
\end{equation}
Gauge transformations shuffle a scale invariant operator quantity
between the two operators $J_{\mu 5}^{\rm con}$ and $K_{\mu}$
whilst keeping $J_{\mu 5}^{GI}$ invariant.

The nucleon matrix element of $J_{\mu 5}^{GI}$ is
\begin{equation}
\langle p,s|J^{GI}_{5 \mu}|p',s'\rangle 
= 2M \biggl[ {\tilde s}_\mu G_A (l^2) + l_\mu l.{\tilde s} G_P (l^2) \biggr]
\label{e8}
\end{equation}
where $l_{\mu} = (p'-p)_{\mu}$
and
${\tilde s}_{\mu} 
= {\overline u}_{(p,s)} \gamma_{\mu} \gamma_5 u_{(p',s')} / 2M $.
Since $J^{GI}_{5 \mu}$ does not couple to a massless 
Goldstone
boson it follows that $G_A(l^2)$ and $G_P(l^2)$ contain
no massless pole terms.
The forward matrix element of $J^{GI}_{5 \mu}$ is well
defined and
\begin{equation}
g_A^{(0)}|_{\rm inv} = E(\alpha_s) G_A (0).
\label{e9}
\end{equation}

We would like to isolate the gluonic contribution to $G_A (0)$
associated with $K_{\mu}$ and thus write $g_A^{(0)}$ 
as the sum of 
(measurable)
``quark'' and ``gluonic'' contributions.
Here one has to be careful because of the gauge dependence of
the operator $K_{\mu}$.
To understand the gluonic contributions to $g_A^{(0)}$ it is
helpful to go back to the deep inelastic cross-section in Section 2.

\subsection{The anomaly and the first moment of $g_1$}

We specialise to the target rest frame and let $E$ denote the 
energy of the incident charged lepton
which is scattered through an angle $\theta$ 
to emerge in the final state with energy $E'$.
Let $\uparrow \downarrow$ denote the longitudinal polarization of
the beam
and $\Uparrow \Downarrow$ denote a longitudinally polarized proton
target.
The spin dependent part of the differential cross-sections is:
\begin{equation}
\Biggl(
{d^2 \sigma \uparrow \Uparrow \over d\Omega dE^{'} } -
{d^2 \sigma \uparrow \Downarrow \over d\Omega dE^{'} } 
\Biggr)
=
{4 \alpha^2 E^{'} \over Q^2 E \nu }
\biggl[ (E+E^{'} \cos \theta ) \ g_1 (x, Q^2) - {2 x M}
\ g_2 (x, Q^2) \biggr] 
\end{equation}
which is obtained from the product of the lepton and hadron
tensors:
\begin{equation}
{d^2 \sigma \over d\Omega dE'} 
= {\alpha^2 \over Q^4} {E' \over E} \ L_{\mu \nu}^A \ W^{\mu \nu}_A
\label{e10}
\end{equation}
Here the lepton tensor
\begin{equation}
L_{\mu \nu}^A = 2 i \epsilon_{\mu \nu \alpha \beta} k^{\alpha} q^{\beta}
\label{e11}
\end{equation}
describes the lepton-photon vertex and 
the hadronic tensor
\begin{equation}
{1 \over M} W^{\mu \nu}_A =
i \epsilon^{\mu \nu \rho \sigma} q_{\rho} 
\biggl(
s_{\sigma} {1 \over p.q} g_1 (x,Q^2)
+ [ p.q s_{\sigma} - s.q p_{\sigma} ] {1 \over M^2 p.q} g_2 (x,Q^2) \biggr)
\label{e12}
\end{equation}
describes the photon-nucleon interaction.

Deep inelastic scattering involves the Bjorken limit:
$Q^2 = - q^2$ and $p.q = M\nu$ both $\rightarrow \infty$ 
with 
$x = {Q^2 \over 2 M \nu}$ held fixed.
In terms of light-cone coordinates this corresponds to taking
$q_- \rightarrow \infty$ 
with 
$q_+ = -x p_+$ held finite.
The leading term in $W_A^{\mu \nu}$ 
is obtained by taking the Lorentz index of $s_{\sigma}$
as $\sigma = +$.
(Other terms are suppressed by powers of ${1 \over q_-}$.)

The flavour-singlet axial charge which is measured in the 
first moment of $g_1$ is 
given by the matrix element
\begin{equation}
2M s_\mu g_A^{(0)} 
= 
\langle p, s| J^{GI}_{\mu5} |p, s \rangle
\nonumber
\end{equation}
If we wish to understand the first moment of $g_1$ in terms 
of the 
matrix elements of anomalous currents
($J_{\mu 5}^{\rm con}$ and $K_{\mu}$),
then we have to understand
the forward matrix element of $K_+$.

Here we are fortunate in that the parton model is formulated in the 
light-cone gauge ($A_+=0$) where the forward matrix elements of $K_+$ 
are invariant.
In the light-cone gauge the non-abelian three-gluon part of $K_+$ 
vanishes. The forward matrix elements of $K_+$ are then invariant 
under all residual gauge degrees of freedom.
Furthermore, 
in this gauge, $K_+$ measures the gluonic ``spin'' content of the 
polarized target \cite{jafpl}
\footnote{Strictly speaking, up to a non-perturbative surface term 
          in the light-cone correlation function.}
.
We find \cite{ccm,etar}
\begin{equation}
G_A^{(\rm A_+ = 0)}(0) = \sum_q \Delta q_{\rm con} 
- f {\alpha_s \over 2 \pi} \Delta g
\label{e13}
\end{equation}
where
$\Delta q_{\rm con}$ is measured by the partially conserved current
$J_{+5}^{\rm con}$
and 
$- {\alpha_s \over 2 \pi} \Delta g$ is measured by $K_+$.
The gluonic term in Eq.(\ref{e13}) 
offers a 
possible source for any OZI violation in $g_A^{(0)}|_{\rm inv}$.
\footnote{Note that non-forward matrix elements of $K_+$ are not 
invariant under residual gauge degrees of freedom 
even in perturbation theory. 
It follows that any extension of this formalism to non-forward 
parton distributions is non-trivial \cite{nfpd}.}

What is the relation between the formal decomposition in Eq.(\ref{e13}) 
and our previous (more physical) expression 
(\ref{eqa3}) ?

\subsection{Questions of gauge invariance}

In perturbative QCD $\Delta q_{\rm con}$ is identified 
with 
$\Delta q_{\rm partons}$ and 
$\Delta g$ is identified 
with $\Delta g_{\rm partons}$
-- see Section 6 and \cite{ccm,bint,etar}.
If we were to work only in the light-cone gauge we might think 
that we have a complete parton model description of the first 
moment of $g_1$.
However, one is free to work in any gauge including a covariant
gauge where the forward matrix elements of $K_+$ 
are not necessarily invariant under the residual gauge degrees 
of freedom \cite{jaffem}.

We illustrate this by an example in covariant gauge.

The matrix elements of $K_{\mu}$ need to be specified with
respect to a specific gauge.
In a covariant gauge we can write
\begin{equation}
\langle p,s|K_\mu |p',s'\rangle
= 2M \biggl[ {\tilde s}_\mu K_A(l^2) + l_\mu l.{\tilde s} K_P(l^2) \biggr]
\label{e14}
\end{equation}
where $K_P$ contains a massless Kogut-Susskind pole \cite{kogut}.
This massless pole cancels with a corresponding massless
pole term in $(G_P - K_P)$.
In an axial gauge $n.A=0$ the matrix elements of the gauge dependent 
operator $K_{\mu}$ will also contain terms proportional to the gauge
fixing vector $n_{\mu}$.

We may define a gauge-invariant form-factor $\chi^{g}(l^2)$
for the topological charge density (\ref{e5}) 
in the divergence of 
$K_{\mu}$:
\begin{equation}
2M \ l.{\tilde s} \ \chi^g(l^2) =
\langle p,s | {g^2 \over 8 \pi^2} G_{\mu \nu} {\tilde G}^{\mu \nu}
 | p', s' \rangle.
\label{e15}
\end{equation}
Working in a covariant gauge, we find
\begin{equation}
\chi^{g}(l^2) = K_A(l^2) + l^2 K_P(l^2)
\label{e16}
\end{equation}
by contracting Eq.(\ref{e14}) with $l^{\mu}$.

When we make a gauge transformation any change 
$\delta_{\rm gt}$
in $K_A(0)$ is compensated
by a corresponding change in the residue of the Kogut-Susskind
pole in $K_P$, viz.
\begin{equation}
\delta_{\rm gt} [ K_A(0) ]
+ \lim_{l^2 \rightarrow 0} \delta_{\rm gt} [ l^2 K_P(l^2) ] = 0.
\label{e17}
\end{equation}
The Kogut-Susskind pole corresponds to the Goldstone 
boson associated with spontaneously broken $U_A(1)$ symmetry
\cite{rjc}.
There is no Kogut-Susskind pole in perturbative QCD.
It follows that the quantity which is shuffled 
between the $J_{+5}^{\rm con}$ and $K_+$ 
contributions to $g_A^{(0)}$ is strictly non-perturbative;
it vanishes in perturbative QCD and is not present in the QCD
parton model.

One can show \cite{jaffem,cron} that the forward matrix elements of 
$K_{\mu}$ are invariant under ``small'' gauge transformations
(which are topologically deformable to the identity) 
but not invariant under ``large'' gauge transformations which 
change the topological winding number.
Perturbative QCD involves only ``small'' gauge transformations;
``large'' gauge transformations involve strictly non-perturbative physics.
The second term on the right hand side of Eq.(\ref{e7}) is a total derivative; 
its matrix elements vanish in the forward direction.
The third term on the right hand side of Eq.(\ref{e7}) 
is associated with the gluon topology \cite{cron}.

The topological winding number is determined by the gluonic boundary 
conditions at 
``infinity'' 
\footnote
{A large surface with boundary which is spacelike with respect 
 to the positions $z_k$ of any operators or fields in the physical
 problem.}
\cite{rjc}.
It is insensitive to local deformations of the gluon 
field $A_{\mu}(z)$ or of the gauge transformation $U(z)$.
When we take the Fourier transform to momentum space 
the topological structure induces a light-cone zero-mode which 
can contribute to $g_1$ only at $x=0$.
Hence, we are led
to consider the possibility that there may be a 
term in $g_1$ which is proportional to $\delta(x)$ \cite{topology}.

It remains an open question whether the net non-perturbative
quantity which
is shuffled between $K_A(0)$ and $(G_A - K_A)(0)$ under ``large''
gauge transformations
is finite or not.
If it is finite and, therefore, physical, then, when we choose
$A_+ =0$,
this non-perturbative quantity must be contained in 
some combination of the $\Delta q_{\rm con}$ and $\Delta g$ in Eq.(\ref{e13}).

Previously, in Sections 3-4, we found that a $J=1$ fixed pole in the real 
part of $A_1$ in the forward Compton amplitude
could also induce a ``$\delta (x)$ correction''
to the sum rule for the first moment of $g_1$ 
through a subtraction at infinity in the dispersion relation
(\ref{eqcl}).
Both the topological $x=0$ term and the subtraction constant
${Q^ 2 \over 2M^2} \beta_1 (Q^2)$ 
(if finite) 
give real coefficients of
${1 \over x}$ terms in Eq.(\ref{eqcm}).
It seems reasonable therefore to conjecture that the physics of 
gluon topology
may induce a $J=1$ fixed pole correction to the Ellis-Jaffe sum rule.

Instantons provide an example how to generate topological $x=0$ 
polarization \cite{topology}.
Quarks instanton interactions flip chirality, thus connecting
left and right handed quarks.
Whether instantons spontaneously or explicitly break axial U(1) 
symmetry depends on the role of zero modes in the quark instanton
interaction and how one should include non local structure in the 
local anomalous Ward identity.
Topological $x=0$ polarization is natural in theories of spontaneous
axial U(1) symmetry breaking by instantons \cite{rjc}
where any instanton induced suppression of $g_A^{(0)}|_{\rm pDIS}$ 
is compensated by a shift of flavour-singlet 
axial charge from quarks carrying finite momentum to a zero mode ($x=0$).
It is not generated by mechanisms \cite{thooft} of explicit U(1) 
symmetry breaking by instantons.

The relationship between the spin structure of the proton and dynamical 
axial U(1) symmetry breaking is further highlighted through 
the flavour-singlet Goldberger-Treiman relation \cite{venez}
which relates $g_A^{(0)}$ to the product of 
the nucleon coupling of the flavour-singlet Goldstone boson 
that would exist in a gedanken world where OZI is exact and 
the first derivative of the QCD topological susceptibility.
The role of the topological charge density in 
low-energy hadron interactions is reviewed in \cite{upps}.
Anomalous glue may play a key role in the structure of the 
light mass (about 1400-1600 MeV) exotic mesons with quantum numbers 
$J^{PC} = 1^{-+}$ 
that have been observed in experiments at BNL and CERN.
These states might be dynamically generated resonances in $\eta' \pi$
rescattering \cite{bassmarco}
(mediated by the OZI violating coupling of the $\eta'$).

\section{Partons and $g_1$}

We now discuss polarized photon gluon fusion, its relation to the axial
anomaly, and importance to semi-inclusive measurements of polarized 
deep inelastic scattering which aim to disentangle the spin-flavour 
structure of the nucleon's sea.

\subsection{Photon gluon fusion}

Consider the polarized photon-gluon fusion process
$\gamma^* g \rightarrow q {\bar q}$.
We evaluate the $g_1$ spin structure function for this process as a
function of the transverse momentum squared of the struck quark, $k_t^2$,
with respect to the photon-gluon direction.
We use $q$ and $p$ to denote the photon and gluon momenta and
use the cut-off $k_t^2 \geq \lambda^2$ to separate the total
phase space into ``hard'' ($k_t^2 \geq \lambda^2$) and ``soft''
($k_t^2 < \lambda^2$) contributions.
One finds \cite{bassbs}:
\begin{eqnarray}
& & 
g_1^{(\gamma^* g)} 
|_{\rm hard} 
=
\nonumber \\
& &
\ \ \
-{\alpha_s \over 2 \pi } {\cut \over \cxx} \Biggl[ (2x-1)(\cx) 
\nonumber \\
& & 
\ \ \ 
\biggl\{
1 - {1 \over {\cut \sqrt{\cxx} }}
\ln \biggl({ {1+\sqrt{\cxx} \cut}\over {1-\sqrt{\cxx} \cut}}
\biggr) \biggr\} 
\nonumber \\
& & 
\ \ \ \ \ \ \ \ \ \ \ \ \ \ 
+ (x-1+{{x P^{2}}\over{Q^{2}}})
{{\left( 2m^{2}(\cxx)- P^{2}x(2x-1)(\cx)\right)}
\over {(m^{2} + \lambda^2) (\cxx) - P^{2}x(x-1+{{x P^{2}}\over{Q^{2}}})}}
\Biggr]
\nonumber \\
\label{eq7}
\end{eqnarray}
for each flavour of quark liberated into the final state.
Here $m$ is the quark mass,
$Q^2 =-q^2$ is the virtuality of the hard photon,
$P^2=-p^2$ is the virtuality of the gluon target,
$x$ is the Bjorken variable ($x= {Q^2 \over 2 p.q}$)
and
$s$ is the centre of mass energy squared,
$
s= (p+q)^2 = Q^2 \bigl( {1 - x \over x} \bigr) - P^2
$,
for the photon-gluon collision.

When $Q^2 \rightarrow \infty$
the expression for $g_1^{(\gamma^* g)}|_{\rm hard}$
simplifies to the leading twist (=2) contribution:
\begin{eqnarray}
g_1^{(\gamma^* g)}|_{\rm hard}
= {\alpha_s \over 2 \pi} \Biggl[ (2x-1) \Biggl\{
\ln {1-x \over x} - 1
+ \ln {Q^2 \over {x(1-x) P^2 + (m^2 + \lambda^2)} } \Biggr\}
& &
\nonumber \\
(1 -x) { {2m^2 - P^2x(2x-1)} \over { m^2 + \lambda^2 - P^2 x(x-1)} }
\Biggr] .
& &
\nonumber \\
\label{eq:11}
\end{eqnarray}
Here we take $\lambda$ to be independent of $x$.
\footnote{
We refer to \cite{bint,bassbs} for a discussion of $x$ dependent cut-offs on 
the virtuality of the struck quark or the invariant mass squared of the 
quark-antiquark pair produced in the photon-gluon collision.
These $x$ dependent cut-offs correspond to different jet definitions and 
different factorization schemes.}
Note that for finite quark masses,
phase space limits Bjorken $x$ to
$x_{max} = Q^2 / (Q^2 + P^2 + 4 (m^2 + \lambda^2))$
and protects
$g_1^{(\gamma^* g)}|_{\rm hard}$
from reaching the $\ln (1-x)$ singularity in Eq. (\ref{eq:11}).
For this photon-gluon fusion process,
the first moment of the ``hard'' contribution is:
\begin{equation}
\int_0^1 dx g_1^{(\gamma^{*} g)}|_{\rm hard}
= - {\alpha_s \over 2 \pi}
\left[1 + \frac{2m^2}{P^2}
\frac{1}{\sqrt{ 1 + {4 (m^2 + \lambda^2) \over P^2} } }
\ln \left(
\frac{\sqrt{1 + {4 (m^2+\lambda^2) \over P^2} -1} }
{\sqrt{1 + {4 (m^2+\lambda^2) \over P^2} +1} } \right) \right]
\label{eq:12}
\end{equation}
The ``soft'' contribution to the first moment of $g_1$ is then
obtained by subtracting Eq. (\ref{eq:12})
from the inclusive first moment (obtained by setting $\lambda =0$).

For fixed gluon virtuality $P^2$ the photon-gluon fusion process
induces two distinct contributions to the first moment of $g_1$.
Consider the leading twist contribution, Eq. (\ref{eq:12}).
The first term, $-{\alpha_s \over 2 \pi}$, in Eq.(\ref{eq:12})
is mass-independent and comes from the region of phase space
where the struck quark carries large transverse momentum squared
$k_t^2 \sim Q^2$.
It measures a contact photon-gluon interaction and is associated
\cite{ccm,bint}
with the
axial anomaly \cite{adler}.
\footnote
{
When we apply the operator product expansion to $g_1^{(\gamma^* g)}$
the first term in Eq.(\ref{eq:12})
corresponds to the gluon matrix element of the anomaly current 
$K_+$ (evaluated in $A_+=0$ gauge).
If we remove the cut-off by setting $\lambda^2$ equal to zero, 
then the second term in Eq.(\ref{eq:12}) 
is the gluon matrix element of $J_{\mu 5}^{\rm con}$ \cite{bint}
}
The second mass-dependent term comes from the region of phase-space
where the struck quark carries transverse momentum $k_t^2 \sim m^2,P^2$.
This 
positive mass dependent term is proportional to the mass
squared of the struck quark.
The mass-dependent in Eq.~(\ref{eq:12})
can safely be neglected for light-quark flavor (up and down) production.
It is very important for strangeness (and charm \cite{bbs,Bass})
production.
For vanishing cut-off ($\lambda^2=0$) this term vanishes in the limit
$m^2 \ll P^2$ and tends to $+{\alpha_s \over 2 \pi}$ when $m^2 \gg P^2$
(so that the first moment of $g_1^{(\gamma^* g)}$
 vanishes in this limit).
The vanishing of
 $\int_0^1 dx g_1^{(\gamma^* g)}$ in the limit $m^2 \ll P^2$
 to leading order in $\alpha_s (Q^2)$
 follows from an application \cite{bassbs} of
 the fundamental Drell-Hearn-Gerasimov sum-rule.

Eq. (\ref{eq:12}) leads to the well known formula quoted in Section 1
\cite{etar,ccm,bint}
\begin{equation}
g_A^{(0)}|_{\rm pDIS} =
\Biggl( \sum_q \Delta q
  - 3 {\alpha_s \over 2 \pi} \Delta g \Biggr)_{\rm partons} 
\label{eq:18}
\end{equation}
(for the non-zero mode contribution to $g_A^{(0)}$)
where $\Delta g$ is the amount of spin carried by polarized gluon
partons in the polarized proton and
$\Delta q_{\rm partons}$ measures
the spin carried by quarks and antiquarks
carrying ``soft'' transverse momentum $k_t^2 \sim m^2, P^2$.
Note that the mass independent contact interaction in Eq.(\ref{eq:12})
is flavour independent.
The mass dependent term associated with low $k_t$ breaks flavour SU(3)
in the perturbative sea.

We next discuss the practical consequence \cite{bass03} of the strange 
quark mass in polarized photon-gluon fusion 
and 
the transverse momentum dependence of the perturbative sea 
generated by photon gluon fusion in semi-inclusive measurements of $g_1$.

\subsection{Sea polarization and semi-inclusive polarized deep inelastic 
scattering}

Semi-inclusive measurements of fast pions and kaons in the current
fragmentation region with final state particle identification can
be used to reconstruct the individual up, down and strange quark
contributions to the proton's spin \cite{close,closem}.
In contrast to inclusive polarized deep inelastic scattering
where the $g_1$ structure function is deduced by detecting only
the scattered lepton, the detected particles in the semi-inclusive
experiments are high-energy (greater than 20\% of the energy of the
incident photon)
charged pions and kaons in coincidence with the scattered lepton.
For large energy fraction $z=E_h/E_{\gamma} \rightarrow 1$
the most probable occurence is that the detected $\pi^{\pm}$ and
$K^{\pm}$
contain the struck quark or antiquark in their valence Fock state.
They therefore act as a tag of the flavour of the struck quark.

New semi-inclusive data reported by the HERMES experiment 
\cite{hermessemi} 
(following earlier work by SMC \cite{smcsemi})
suggest that the light-flavoured (up and down) sea measured in these
semi-inclusive experiments contributes close to zero to the proton's
spin.
For the region $0.023 < x < 0.3$
the extracted $\Delta s$
integrates to the value $+0.03 \pm 0.03 \pm 0.01$ 
which contrasts with the negative value for the polarized 
strangeness 
(\ref{eqa2}) extracted from inclusive measurements of $g_1$.

An important issue for semi-inclusive measurements is the angular 
coverage of the detector \cite{bass03}.
The non-valence spin-flavour structure of the proton extracted 
from semi-inclusive measurements of polarized deep inelastic scattering
may depend strongly on the transverse momentum (and angular)
acceptance of the detected final-state hadrons which are used 
to determine the individual polarized sea distributions.
The present semi-inclusive experiments detect final-state hadrons
produced only at small angles from the incident lepton beam 
(about 150 mrad angular coverage) 
whereas
the perturbative QCD
``polarized gluon interpretation'' 
\cite{etar} of
the inclusive measurement (\ref{eqa2}) 
involves physics at
the maximum transverse momentum \cite{ccm,bass03} and large angles.

\begin{figure}[h]
\includegraphics{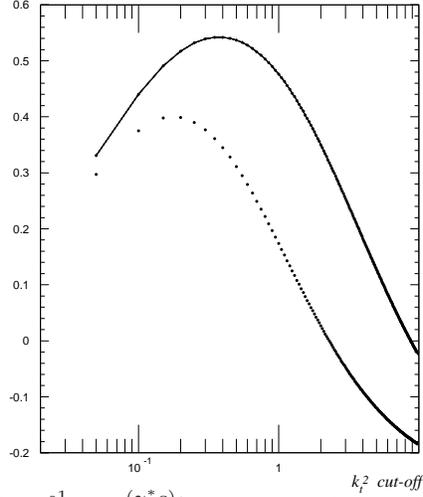} 
\begin{center}
\vspace{6.5cm}
\parbox{12.0cm}
{\caption[Delta]
{
$\int_0^1 dx \ g_1^{(\gamma^* g)}|_{\rm soft}$
for polarized strangeness production
with $k_t^2 < \lambda^2$
in units of ${\alpha_s \over 2 \pi}$.
Here
$Q^2=2.5$GeV$^2$ (dotted line) and 10GeV$^2$ (solid line).
}
\label{fig3}}
\end{center}
\end{figure}
\begin{figure}[h]
\includegraphics{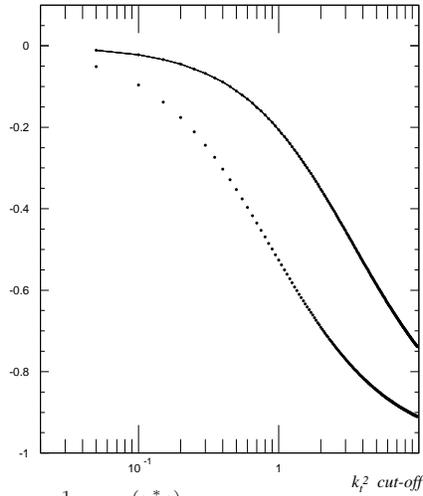} 
\begin{center}
\vspace{6.5cm}
\parbox{12.0cm}
{\caption[Delta]
{
$\int_0^1 dx \ g_1^{(\gamma^* g)}|_{\rm soft}$
for light-flavor ($u$ or $d$) production
with $k_t^2 < \lambda^2$
in units of ${\alpha_s \over 2 \pi}$.
Here
$Q^2=2.5$GeV$^2$ (dotted line) and 10GeV$^2$ (solid line).
}
\label{fig4}}
\end{center}
\end{figure}

Let
$g_1^{(\gamma^* g)}|_{\rm soft} (\lambda)$ denote the contribution
to
$g_1^{(\gamma^* g)}$ for photon-gluon fusion where the hard photon
scatters on the struck quark or antiquark carrying
transverse momentum $k_t^2 < \lambda^2$.
Figs. 3 and 4 show the first moment of $g_1^{(\gamma^* g)}|_{\rm soft}$
for the strange and light (up and down) flavour production
respectively as a function of the transverse momentum cut-off $\lambda^2$.
Here we set $Q^2 =2.5$GeV$^2$
(corresponding to the HERMES experiment) and 10GeV$^2$ (SMC).
Following \cite{ccm},
we take $P^2 \sim \Lambda_{\rm qcd}^2$ and set $P^2 = 0.1$GeV$^2$.
Observe the small value for the light-quark
sea polarization at low transverse momentum and
the positive value for the integrated
strange sea polarization at low $k_t^2$:
$k_t < 1.5$GeV at the HERMES $Q^2=2.5$GeV$^2$.
When we relax the cut-off, increasing the acceptance of the experiment,
the measured strange sea polarization changes sign and becomes
negative (the result implied by fully inclusive deep inelastic measurements).
Note that for $\gamma^*g$ fusion the cut-off $k_t^2 < \lambda^2$
is equivalent to a cut-off on the angular acceptance
$\sin^2 \theta < 4 \lambda^2 / \{s - 4m^2 \}$
where $\theta$ is defined relative to the photon-gluon direction
and $s$ is the centre of mass energy for the photon-gluon collision.
Leading-twist negative sea polarization at
$k_t^2 \sim Q^2$ corresponds, in part,
to final state hadrons produced at large angles.
For HERMES the average transverse momentum of the detected
final-state fast
hadrons is less than about 0.5 GeV whereas for SMC
the $k_t$ of the detected fast pions was less than about 1 GeV.
New semi-inclusive measurements with increased luminosity and a
$4 \pi$ detector, as proposed
for the next generation Electron Ion Collider facility
in the United States, would be extremely useful to map out
the transverse momentum distribution of the total polarized
strangeness (\ref{eqa2}) measured in inclusive deep inelastic scattering.

\section{The main issues}

\begin{itemize}
\item
Are there fixed pole corrections to spin sum rules for polarized
photon nucleon scattering ? 
If yes, which ones ?
\item
How large is the gluon polarization in the proton ?
\item
Is gluon topology important in the spin structure of the proton ?
\item
What happens to ``spin'' 
in the transition from current to constituent
quarks through dynamical axial U(1) symmetry breaking ?
\item
What is the $x$ and $k_t$ dependence of the (negative) polarized
strangeness extracted from inclusive polarized deep inelastic scattering ?
\item
How do the effective intercepts for small $x$ physics change in the 
transition region between polarized photoproduction and polarized
deep inelastic scattering ?
\end{itemize}

\vspace{1.0cm}

{\bf Acknowledgements:} \\

I thank R. L. Jaffe, Z.-E. Meziani and A. W. Thomas for helpful 
discussions,
and A. Bialas and M. Praszalowicz 
for creating a most stimulating scientific meeting and for excellent 
hospitality.
SDB is supported by a Lise Meitner Fellowship (M683 and M770) of the 
Austrian Science Fund (FWF).

\vspace{1.0cm}


\end{document}